\documentclass[journal]{IEEEtran}

\pdfoutput=1
\usepackage[pdftex]{graphicx}
\graphicspath{{./Figures}}
\usepackage{empheq}
\usepackage{amssymb,amsmath}
\usepackage{epstopdf}
\usepackage{float}
\usepackage{subcaption}
\usepackage{algpseudocode}
\usepackage{verbatim}
\usepackage[flushleft]{threeparttable}
\usepackage{booktabs}
\usepackage{amsthm}
\newtheorem{remark}{Remark}

\usepackage[hang,flushmargin]{footmisc} 

\usepackage[ruled, lined, ,longend, linesnumbered]{algorithm2e}
\SetKwInOut{Require}{Require}
\SetKw{KwTo}{to}

\usepackage{xcolor}
\usepackage{hyperref}
\hypersetup{%
    colorlinks=true,
    linkcolor={blue!50!black},
    citecolor={blue!50!black},
    urlcolor={blue!50!black}
}

\usepackage{fancyhdr}

\def\vv#1{{ \rm \bf{#1}}} 
\def\R{\mathbb{R}} 
\def\Z{\mathbb{Z}} 
\newcommand{\T}{^\top} 
\def\diag{\texttt{diag}} 

\begin{document}
\pagestyle{fancy}
\title{\LARGE Efficient online update of model predictive control in embedded systems using first-order methods}
\author{Victor~Gracia,~Pablo~Krupa,~Teodoro~Alamo,~Daniel~Limon%
    \thanks{Department of Systems Engineering and Automation, Universidad de Sevilla, Spain (e-mails: \texttt{vgracia@us.es}, \texttt{pkrupa@us.es}, \texttt{talamo@us.es}, \texttt{dlm@us.es}). Corresponding author: Victor Gracia.}%
    \thanks{The authors acknowledge support from Grants PID2022-141159OB-I00 and PID2022-142946NA-I00, funded by MCIN/AEI/ 10.13039/501100011033 and by “ERDF A way of making Europe”, and Grant PDC2021-121120-C21 funded by MCIN/AEI/10.13039/501100011033 and by the “European Union NextGenerationEU/PRTR”.
    }
}

\maketitle
\thispagestyle{fancy}

\begin{abstract}
    
Model Predictive Control (MPC) is typically characterized for being computationally demanding, as it requires solving optimization problems online; a particularly relevant point when considering its implementation in embedded systems.
To reduce the computational burden of the optimization algorithm, most solvers perform as many offline operations as possible, typically performing the computation and factorization of its expensive matrices offline and then storing them in the embedded system.
This improves the efficiency of the solver, with the disadvantage that online changes on some of the ingredients of the MPC formulation require performing these expensive computations online.
This article presents an efficient algorithm for the factorization of the key matrix used in several first-order optimization methods applied to linear MPC formulations, allowing its prediction model and cost function matrices to be updated online at the expense of a small computational cost.
We show results comparing the proposed approach with other solvers from the literature applied to a linear time-varying system.
\end{abstract}

\section{Introduction} \label{sec:introduction}

Model Predictive Control (MPC) \cite{camacho2013model, rawlings_model_2017} is a well-known control policy due to its ability to optimize the system operation while satisfying its constraints.
However, its control law is computationally demanding, as it requires solving an optimization problem every sample time.
Many linear MPC solvers have been proposed in the literature using different optimization methods, such as active set \cite{9779534}, interior point \cite{domahidi}, or various first-order methods (FOM) such as gradient descent \cite{9810004}, Douglas-Rachford \cite{7419888}, the alternating direction method of multipliers (ADMM) \cite{MAL-016} or FISTA \cite{doi:10.1137/080716542}, to name a few.
In particular, FOMs are a popular choice for their use in embedded systems \cite{GISELSSON20142303, 7330932}, due to their simplicity, good practical performance, and the availability of certification results on their iteration complexity \cite{6315076, Krupa_TCST_21, krupa2023certification}.
Due to the low-resource nature of embedded systems, the implementation of these methods often avoids any non-strictly necessary online computations by computing all possible solver components offline, which are then stored in the embedded system \cite{Krupa_TCST_21, osqp, scs}.
This approach is particularly advantageous when the computation of these elements requires computationally expensive operations, such as matrix inversion or factorization.
In particular, factorization of the key solver ingredients is an approach typically taken to reduce the computational complexity per-iteration of the algorithm, see, e.g., \cite{domahidi, Krupa_TCST_21, osqp, Garstka_2021}.
The disadvantage, however, is that in a practical setting it may no longer be possible to change online some of the ingredients of the MPC controller, such as its prediction model or cost function matrices. 
That is, the operations required to update the solver for the new resulting optimization problem may be too computationally expensive for the embedded system given the sample time, mostly due to the previously mentioned matrix computations and factorization.

A typical scenario that requires online adjustment of some of these elements is when dealing with linear time-varying (LTV) or linear parameter-varying (LPV) systems \cite{BOKOR200512}, where the prediction model of the MPC controller is typically updated every sample time.
In fact, the need to update the prediction model will be a requirement when implementing any adaptive MPC control scheme \cite{TAO20142737}.
Additionally, being able to tune online the cost function matrices of the MPC controller is a welcome feature when working in a practical setting, either to fine tune the controller online to improve its performance when controlling the real system or to adapt to a changing economic criterion on the ``best" way to control it.

Some quadratic programming (QP) solvers provide the option to recompute and refactor its computationally expensive ingredients online, e.g., \cite{osqp, scs}.
However, the required computation time can be significant when compared to their solve time.
In \cite{6571216}, a Ricatti recursion is used to compute the factorization of the linear system to be solved at each iteration of the algorithm. The computation time of this factorization procedure does not depend on the prediction horizon of the MPC controller if its terminal cost is taken as the solution of the discrete-time Ricatti equation associated to the infinite-horizon unconstrained problem.


In this article, we present an algorithm for updating the ingredients used in several FOM-based solvers, including the popular methods ADMM and FISTA, when applied to standard linear MPC formulations whose prediction model and/or weight matrices change online.
This algorithm, in essence, is an efficient computation of the Cholesky decomposition of the main computationally expensive matrix used in several solvers \cite{Krupa_TCST_21, 9632418}.
The key feature of the proposed approach is that the matrix factorization is performed exploiting the particular structure that arises from many standard linear MPC formulations, providing a computational advantage when compared to the online factorization routines provided by solvers for generic QPs from the literature.
This may lead to tractable re-factorization of the solver ingredients in cases where the re-factorization schemes of generic QP solvers are too slow due to the short sample time or the limitations of the embedded system where the solver is implemented.
To illustrate this, we show numerical results comparing the proposed approach to several state-of-the-art QP solvers applied to an LTV system.

\newpage The article is structured as follows.
Section~\ref{sec:problem} presents the problem formulation.
In Section~\ref{sec:online}, we present the proposed algorithm for the online computation and factorization of the solver ingredient.
Section~\ref{sec:numerical} shows the numerical results.
Finally, Section~\ref{sec:conclusions} summarizes the main conclusions.

\subsubsection*{Notation}
The symbol $\Z_a^b$ denotes the set of integers from the integers $a$ to $b$, both included.
For a matrix $A$, its element in row $i$ and column $j$ is denoted as  $A_{i,j}$.
We denote $\| x \|_{Q} \doteq \sqrt{x\T Q x}$.
The symbol $I_{n}$ denotes the identity matrix of dimension $n$.
Given two vectors $x$ and $y$, $x \leq (\geq) \; y$ denotes componentwise inequalities.
For vectors $x_1$ to $x_N$, $(x_{1}, x_{2}, \dots, x_{N})$ denotes the column vector formed by their concatenation.
Given scalars and/or matrices $A_1, A_2, \dots, A_N$ (not necessarily of the same dimensions), we denote by $\diag(A_1, A_2, \dots, A_N)$ the block diagonal matrix formed by the concatenation of $A_1$ to $A_N$.

\section{Problem formulation} \label{sec:problem}

Consider a discrete-time system described by
\begin{equation}\label{pred_model}
    x(k+1) = Ax(k) + Bu(k),
\end{equation}
where $x(k) \in \R^n$ and $u(k) \in \R^m$ are the state and input at sample time $k$, respectively, subject to box constraints
\begin{subequations} \label{box_limits}
    \begin{align}
        \underline{x}(k) & \leq  x(k) \leq \overline{x}(k),\\
        \underline{u}(k) & \leq  u(k) \leq \overline{u}(k),
    \end{align}
\end{subequations}
where the bounds $\underline{x}(k)$, $\overline{x}(k) \in \R^n$ and $\underline{u}(k)$, $\overline{u}(k) \in \R^m$ satisfy $\underline{x}(k) < \overline{x}(k)$ and $\underline{u}(k) < \overline{u}(k)$. 
The control objective is to steer the system to a reference pair $(x_r,u_r)$, assumed to be an admissible equilibrium point of the system.

This objective can be attained using one of many linear MPC formulations proposed in the literature \cite{camacho2013model, rawlings_model_2017, Krupa_TCST_21}.
For instance, consider the standard linear MPC formulation
\begin{subequations} \label{laxMPC}
    \begin{align}
        \min_{\vv{x}, \vv{u}} \;& \sum_{j=0}^{N-1} \left( ||x_{j}{-}x_{r}||_{Q}^2 + ||u_{j}{-}u_{r}||_{R}^2 \right) + ||x_{N}{-}x_{r}||_T^2\\
        \rm s.t. \;&  x_{j+1}= Ax_j + Bu_j, \, j \in \Z_0^{N-1},\\ 
        & x_{0}=x(k),\\
        & \underline{x}_{j} \leq  x_{j} \leq \overline{x}_{j}, \, j \in \Z_1^{N}, \label{laxMPC:box:x} \\ 
        & \underline{u}_{j} \leq  u_{j} \leq \overline{u}_{j}, \, j \in \Z_0^{N-1}, \label{laxMPC:box:u}
    \end{align}
\end{subequations}
where $N$ is the prediction horizon; $Q \in \R^{n \times n}$, $R \in \R^{m \times m}$ and $T \in \R^{n \times n}$ are positive definite matrices; and the decision variables $\vv{x}$ and $\vv{u}$ contain the state and input variables, respectively, along the prediction horizon.

It is well known that, under mild assumptions, system \eqref{pred_model} under the control law $u(k) = u^*_0$, where $u_0^*$ is the optimal solution of the decision variable $u_0$ of \eqref{laxMPC}, asymptotically converges to the admissible reference $(x_r, u_r)$ without violating the system constraints \cite{rawlings_model_2017}.
Therefore, the implementation of the MPC controller requires obtaining the optimal solution of \eqref{laxMPC} at each sample time.

We remark that formulation \eqref{laxMPC} has been selected as an example instead of any other linear MPC formulation in order to provide the reader with concrete expressions. However, as shown later on, the proposed update procedure can be adapted to other linear MPC formulations with minor modifications. 

Problem \eqref{laxMPC} can be rewritten as a standard QP problem 
\begin{subequations}\label{canonical_qp}
    \begin{align}
        \min_{z}  \;& \frac{1}{2} z\T H z + q\T z \\
        {\rm s.t.}  \;& Gz= b,\\
                    & \underline{z} \leq z \leq \overline{z},
    \end{align}
\end{subequations}
by taking
\begin{subequations}\label{canonical_ingredients}
    \begin{align}
        &H = \diag(R, Q, R, \dots, Q, R, T), \label{canonical_ingredients:H} \\
        &q = -(R u_r, Q x_r, R u_r, \dots, Q x_r, R u_r, T x_r),\\
                &G =
                \begin{bmatrix}
                    B & -I_{n} & 0 & \cdots & \cdots & 0\\
                    0 & A & B & -I_{n} & \cdots & 0\\
                    0 & 0 & \ddots & \ddots & \ddots & 0\\
                    0 & 0 & 0 & A & B & -I_{n}
                \end{bmatrix}, \label{canonical_ingredients:G} \\
        &b = -(Ax(k), 0, 0, ... , 0),\\
        &\underline{z} = (\underline{u}_0, \underline{x}_1, \underline{u}_1, ... , \underline{x}_{N-1}, \underline{u}_{N-1}, \underline{x}_N), \\
        &\overline{z} = (\overline{u}_0, \overline{x}_1, \overline{u}_1, ... , \overline{x}_{N-1}, \overline{u}_{N-1}, \overline{x}_N).
    \end{align}
\end{subequations}

There are many optimization methods that can efficiently solve problem \eqref{canonical_qp} (active-set method, interior point methods, FOMs, etc.).
Among them, FOMs have received a significant amount of attention from the control community, especially when dealing with the implementation of MPC in embedded systems, whose restricted computational and memory resources are a good match for these simple-to-implement algorithms with low resource requirements.
Furthermore, a major advantage of solving \eqref{canonical_qp} using FOMs is that the ingredients of the optimization method typically have simple sparse structures due to the sparse, banded nature of \eqref{canonical_ingredients:H} and \eqref{canonical_ingredients:G}; a fact exploited by several QP solvers from the literature to obtain efficient implementations, e.g., \cite{Krupa_TCST_21, osqp, Garstka_2021}.

For instance, in \cite{Krupa_TCST_21} the authors propose sparse solvers based on ADMM or FISTA for problem \eqref{canonical_qp} arising from simple standard MPC formulations such as \eqref{laxMPC}.
The proposed solvers take advantage of the particular structures of $H$ and $G$ that arise from these MPC formulations, i.e., \eqref{canonical_ingredients:H} and \eqref{canonical_ingredients:G}.
Specifically, they require solving a linear system $W z = d$ at each iteration of the algorithm, where $W \doteq G(H+\rho I)^{-1}G\T$ for some non-negative scalar $\rho$ whose value depends on the specific FOM being used.
The solvers in \cite{Krupa_TCST_21} solve this linear system by backward-forward substitution using the Cholesky decomposition of $W$, i.e., the matrix $W_c$ satisfying $W=W_c\T W_c$, whose simple banded structure is exploited to provide an efficient algorithm.
In particular, the structure of $W_c$ resulting from problem \eqref{canonical_qp} is given by
\begin{equation}\label{alpha_beta}
    W_c =
    \begin{bmatrix}
        \beta^1 & \alpha^1 & 0 & \cdots & 0\\
        0 & \beta^2 & \alpha^2 & \cdots & 0\\
        \vdots & \vdots & \beta^3 & \ddots & \vdots\\
        \vdots & \vdots & \vdots & \ddots & \alpha^{N-1}\\
        0 & 0 & 0 & \cdots & \beta^{N}
    \end{bmatrix},
\end{equation}
where $\beta^{k} \in \R^{n \times n}$, $k \in \Z_1^N$, are upper triangular matrices and $\alpha^{k} \in \R^{n \times n}$, $k \in \Z_1^{N-1}$, are dense.
Matrices $\alpha^k$ and $\beta^k$, which depend on the values of the MPC ingredients $A$, $B$, $Q$, $R$ and $T$, are the main computationally-expensive ingredients of the solver, and are thus typically computed offline and stored in the embedded system, where they are used to efficiently solve the system of equations $W z = d$ \cite[\S 5.1]{krupa2021implementation}.

A limitation of these FOM solvers, therefore, is that changing some of the MPC ingredients online requires recomputing and/or refactoring certain matrices, which can be rather time-consuming.
In the previous example, changing any of the ingredients $A$, $B$, $Q$, $R$ or $T$ would require an online computation of $W_c$ \eqref{alpha_beta}.
In other solvers from the literature, such as \cite{osqp, scs, Garstka_2021}, similar sparse matrices need to be recomputed and refactored if similar MPC ingredients are changed.

The following section presents an efficient algorithm for updating the $\beta^k$ and $\alpha^k$ sub-matrices of $W_c$ \eqref{alpha_beta} obtained from QP problems \eqref{canonical_qp} whose $H$ and $G$ ingredients have banded structures similar to the ones shown in \eqref{canonical_ingredients:H} and \eqref{canonical_ingredients:G}.
The motivation behind this algorithm is to provide optimization solvers which can efficiently deal with MPC ingredients that change online, such as when dealing with LTV or LPV systems instead of the time-invariant system \eqref{pred_model}.
We note that MPC formulation \eqref{laxMPC} is taken as an example whose associated $W_c$ matrix has the structure shown in \eqref{alpha_beta}, but that most other standard linear MPC formulations lead to nearly identical structures, requiring, as we will show, only minor modifications of the procedure presented in the following~section.

\section{Online computation of the solver ingredients} \label{sec:online}

\begin{algorithm}[t]
    \DontPrintSemicolon
    \caption{Computation of the sub-matrices of $W_c$}
    \label{Algorithm1}
    \Require{Igredients that affect $W$ ($A$, $B$, $Q$, $R$, $T$)}
    Compute the matrices in $\Gamma$ \eqref{Gamma_expressions} \label{Algorithm1:matrices}\;
    \For{$k=1$ \KwTo $N-1$}{
        Compute $\beta^{k}$ using \eqref{beta:eq} and \eqref{beta:ineq}\;
        Compute $\alpha^{k}$ using \eqref{alpha_equation}\;
    }
    Compute $\beta^{N}$ using \eqref{beta:eq} and \eqref{beta:ineq} \;
    \KwOut{$\beta^1 \ \text{to} \ \beta^N, \alpha^1 \ \text{to} \ \alpha^{N-1}$}
\end{algorithm}

We now present equations for recursively computing sub-matrices $\alpha^{k}$ and $\beta^{k}$ of $W_c$ \eqref{alpha_beta} associated with the QP problem \eqref{canonical_qp} for the MPC formulation \eqref{laxMPC}.
That is, we provide expressions for computing the non-zero sub-matrices of the Cholesky decomposition of matrix $W \doteq G(H+\rho I)^{-1}G\T$.
We present the final expressions of the equations.
The appendix provides an intuitive overview of how they were obtained.

Let us introduce the notation
\begin{subequations}\label{shortcuts}
	\begin{align}
        &\hat{Q} \doteq (Q+\rho I)^{-1}, \, \hat{R} \doteq (R+\rho I)^{-1}, \, \hat{T} \doteq (T+\rho I)^{-1}, \label{shortcuts:cost} \\
		& Y \doteq A\hat{Q}A\T, \, Z \doteq B\hat{R}B\T,
		\\ & X_Q \doteq Y+Z+\hat{Q}, \; X_T \doteq Y+Z+\hat{T}, \label{X_T}
	\end{align}
\end{subequations}
where we recall that the value of $\rho \geq 0$ depends on the FOM being used \cite{Krupa_TCST_21}.
Setting $\alpha^0 = 0$ for convenience, we denote
\begin{subequations} \label{Gamma_expressions}
	\begin{empheq}[left={\Gamma^k \doteq \empheqlbrace}]{align}
		&Z+\hat{Q}, & \text{if} & \ k=1,  \label{Gamma_expressions:k_1} \\
		&X_Q, &\text{if} & \ k \in \Z_{2}^{N-1}, \label{Gamma_expressions:k}\\
		&X_T, &\text{if} & \ k=N,  \label{Gamma_expressions:k_N}
	\end{empheq}
\end{subequations}
\begin{equation} \label{gamma_expression}
	\gamma^k_{i,j} \doteq \sum_{l=1}^{i-1}\beta_{l,i}^k \beta_{l,j}^k + \sum_{q=1}^{n} \alpha_{q,i}^{k-1} \alpha_{q,j}^{k-1}.
\end{equation}
In \eqref{Gamma_expressions}, $\Gamma^k$ can be seen intuitively as a variable that gathers the elements of the diagonal band of $W$ that affect $\beta^k$. In \eqref{gamma_expression}, $\gamma^k$ groups terms of the non-diagonal elements of $\beta^k$ and elements of $\alpha^{k-1}$ that affect subsequent elements of $\beta^k$.
Keeping this in mind, the nonzero elements $\beta^k_{i,j}$ of $\beta^k \in \R^{n \times n}, \, k \in \Z_{1}^{N}$, and the elements $\alpha^k_{i,j}$ of $\alpha^k \in \R^{n \times n}, \, k \in \Z_{1}^{N-1}$, are given~by
\begin{subequations} \label{lax_formulas}
	\begin{align}
		&\beta^k_{i,i} = \sqrt{\Gamma_{i,i}^k-\gamma^k_{i,i}}, \quad  i \in  \Z_{1}^{n}, \label{beta:eq}\\
        &\beta^k_{i,j} = \frac{\Gamma_{i,j}^k - \gamma^k_{i,j}}{\beta_{i,i}^k}, \quad j \in \Z_{i+1}^n, \label{beta:ineq}\\
        &(\beta^k)\T \alpha^k = - (A \hat{Q})\T. \label{alpha_equation}
	\end{align}
\end{subequations}

Note that the expressions in \eqref{shortcuts:cost} require matrix inversions, which are cheap when $Q, R$ and $T$ are small and/or diagonal.
However, if this is not the case, one could directly provide $\hat{Q}, \hat{R}$ and $\hat{T}$ in order to reduce its computation time.
The computation of $\alpha^k$ using \eqref{alpha_equation} is simple, since each of its columns is computed as the solution of the upper-triangular linear system $(\beta^k)\T \alpha^k_{[j]} = - (A \hat{Q})\T_{[j]}$, where we use subindex $[j]$ to indicate column $j$ of the matrix.

Algorithm~\ref{Algorithm1} shows the procedure for computing $\alpha^k$ and $\beta^k$. The computational complexity of Algorithm \ref{Algorithm1} when applied to the $W$ resulting from \eqref{laxMPC} is $\mathcal{O}(m^3+ n^3 N)$ flops. 

\begin{remark}
    The computational complexity of Algorithm \ref{Algorithm1} can be reduced if certain additional assumptions are made, such as matrices $Q$ and $R$ being diagonal.
\end{remark}

An interesting point about equations \eqref{gamma_expression} and \eqref{lax_formulas} is that they are a consequence of the structure of matrix $W_c$ shown in \eqref{alpha_beta}.
That is, equations \eqref{gamma_expression} and \eqref{lax_formulas} are valid for any MPC formulation whose solver requires solving a linear system $W z = d$ whose Cholesky factorization $W_c$ has the structure shown in~\eqref{alpha_beta}.
Thus, the only difference between Algorithm~\ref{Algorithm1} applied to each formulation is the matrices required to determine the value of $\Gamma^k$ \eqref{Gamma_expressions} and, possibly, the matrix on the right-hand side of \eqref{alpha_equation}.
For the MPC formulation \eqref{laxMPC}, the values of $\Gamma^k$ for each $k$ are shown in \eqref{Gamma_expressions} and \eqref{shortcuts}, but for other MPC formulations these expressions may differ.

Indeed, let us now consider the standard MPC formulation with terminal equality constraint \cite[Eq. (8)]{Krupa_TCST_21}, given by
\begin{subequations}\label{equMPC}
    \begin{align}
        \displaystyle \min_{\textbf{x,u}} & \quad \sum_{j=0}^{N-1} ||x_{j}-x_{r}||_{Q}^2 + ||u_{j}-u_{r}||_{R}^2 \\
        \rm s.t. &  \quad x_{j+1}= Ax_j + Bu_j, j \in \Z_1^{N-1},\\ 
        & \quad x_{0}=x(k),\\
        & \quad \underline{x}_{j} \leq  x_{j} \leq \overline{x}_{j}, \, j \in \Z_1^{N},\\ 
        & \quad \underline{u}_{j} \leq  u_{j} \leq \overline{u}_{j}, \, j \in \Z_0^{N-1},\\
        & \quad x_{N}=x_{r}. \label{terminal_eq_constraint}
    \end{align}
\end{subequations}
In this case, matrices $H$ and $G$ of the resulting QP problem \eqref{canonical_qp} are nearly identical to the ones shown in \eqref{canonical_ingredients:H} and \eqref{canonical_ingredients:G}; the only differences being in the final rows of said matrices due to its different terminal ingredients.
Thus, equations \eqref{Gamma_expressions}-\eqref{lax_formulas} are still valid for updating its $\alpha^k$ and $\beta^k$ matrices, with the exception that \eqref{Gamma_expressions:k_N} now uses $Y+Z$ instead of $X_T$.
This change is a consequence of the inclusion of the terminal equality constraint and the removal of the terminal cost in \eqref{equMPC}, which results in a different expression for $\Gamma^N$.
However, the expressions for the other submatrices of $W_c$ do not change.
For other standard linear MPC formulations whose $H$ and $G$ matrices maintain banded structures similar to the ones shown in \eqref{canonical_ingredients:H} and \eqref{canonical_ingredients:G}, similar small modifications of $\Gamma$ will be required, but overall Algorithm~\ref{Algorithm1} will remain the same.
This is the case, for instance, with the MPC formulations from \cite[\S V.A]{domahidi}, \cite{7330932, Krupa_TCST_21, 6571216}.
Furthermore, we find that this same procedure may also apply to some non-standard MPC formulations, such as in the MPC for tracking solver presented in \cite{9632418} (c.f., \cite[Eq. (11)]{9632418}).

\begin{remark}
The expression of $\Gamma^N$ for \eqref{equMPC} differs from the one for \eqref{laxMPC} because \eqref{terminal_eq_constraint} adds an equality constraint, affecting the structure of \eqref{canonical_ingredients:H} and \eqref{canonical_ingredients:G}, and thus $W$. However, if we consider the more general terminal constraint $x_N \in \mathcal{X}_f$, where $\mathcal{X}_f$ is a polyhedron, inequality constraints are added instead of an inequality constraint.
Therefore, Algorithm~\ref{Algorithm1} would still be valid as long as the inequality constraints are handled so as to not affect the banded structure of $W$ exploited by Algorithm~\ref{Algorithm1}, e.g., uncoupling the equality and inequality constraints using ADMM as in~\cite{Krupa_TCST_21} or by using a suitable dualization approach of the inequality constraints as in~\cite{6571216}.
\end{remark}

\begin{remark} \label{rem:generic}
An alternative approach to the one taken here would be to implement a generic sparse Cholesky decomposition procedure for sparse matrices.
The use of generic sparse matrix factorization routines is used by some QP solvers to provide an efficient way to update the solver ingredients online, such as in the OSQP solver \cite{osqp}, which implements a generic sparse LDL factorization routine.
This approach, while efficient due to its sparse implementation, is for generic sparse matrices, i.e., using generic sparse matrix representations such as the \textit{``compressed sparse column"}.
In contrast, our approach does not make use of sparse matrix representations, but instead directly computes the non-zero elements of the Cholesky decomposition of $W$ by exploiting the specific structure of the matrices $H$ and $G$ shown in \eqref{canonical_ingredients:H} and \eqref{canonical_ingredients:G}.
That is, our approach loses generality in favor of improving its efficiency by being particularized to the specific structure that typically arises in standard linear MPC formulations, thus providing better computation times than the ones obtained using the update procedures of generic QP solvers.
\end{remark}

\begin{remark}
    The expressions provided in \eqref{shortcuts}-\eqref{lax_formulas} for the computation of $\alpha^k,~k \in \mathbb{Z}_1^{N-1}$, and $\beta^k,~k \in \mathbb{Z}_1^N$, can be easily adapted in order to handle variations in $A$, $B$, $Q$ and $R$ between prediction steps, i.e., having $A_j$, $B_j$, $Q_j$, $R_j$, $j \in \mathbb{Z}_1^{N}$, in problem \eqref{laxMPC}. In this case, \eqref{shortcuts} includes the inversion of $N$ matrices, i.e., computing $\hat{Q}^{-1}_j$, $\hat{R}^{-1}_j$, $j \in \mathbb{Z}_1^{N}$, as well as the matrix multiplications required to compute $Y_j = A_j \hat{Q}_j A\T_j$ and $Z_j = B_j\hat{R}_j B\T_j, \ j\in \mathbb{Z}_1^{N}$. Consequently, the number of operations of Algorithm \ref{Algorithm1} increases. Particularly, it presents a computational complexity of $\mathcal{O}( (m^3+n^3)N)$ flops.
\end{remark}

\section{Numerical results} \label{sec:numerical}

\begin{table*}[t]
    \centering
    \small
    \begin{tabular}{lcccccccccccc}   
        & & \multicolumn{4}{c}{Update-time [microseconds]} & \multicolumn{2}{c}{}	& \multicolumn{4}{c}{Solve-time [microseconds]}
        \\	
        \cline{3-6} \cline{9-12}
        Solver  & $N$               & Average         & Median        & Maximum		& Minimum  & & & Average           & Median         & Maximum		& Minimum   & \%       \\ \hline 
        FISTA                & 5                  & 1.03~(0.08)            & 1.02~(0.07)          & 1.30~(0.17)	        & 0.97~(0.06)     & & & 2.69              & 2.06           & 12.25       & 1.36      & 27.75   \\
        ADMM                 & 5                  & 1.05~(0.11)            & 1.04~(0.10)          & 1.36~(0.20)            & 0.96~(0.08)     & & & 71.01             & 23.49          & 686.3       & 15.14      & 1.45   \\ 
        OSQP                 & 5                    & 18.03           & 18.01         & 19.81	        & 17.45    & & & 20.41             & 18.39          & 43.29       & 17.59     & 46.91   \\
        SCS                  & 5                   & 101.6           & 101.6         & 109.9	        & 99.20    & & & 180.1             & 212.2          & 245.7       & 104.1     & 36.08  \\ \hline 
        FISTA               & 15                    & 1.68~(0.20)            & 1.67~(0.19)          & 2.11~(0.57)	        & 1.62~(0.17)     &  & & 6.14             & 4.47           & 29.40       & 2.71      & 21.52  \\
        ADMM                & 15                   & 1.81~(0.22)           & 1.80~(0.21)          & 2.35~(0.47)           & 1.72~(0.19)     & & &  156.5            & 100.2          & 1428.3       & 50.83      & 1.15  \\
        OSQP                & 15                  & 36.95           & 36.93         & 38.73	        & 35.94    & & &  53.39           & 47.54         & 159.65      & 46.05    & 40.90   \\
        SCS                 & 15                    & 233.7           & 233.7         & 244.9	        & 229.6    & & &  560.8           & 597.7         & 663.6      & 339.1    & 29.42   \\
        \hline
    \end{tabular}
\begin{tablenotes}
\item The update times of algorithms FISTA and ADMM have been obtained using SPCIES with Algorithm \ref{Algorithm1} in its update procedure. In brackets, we include the update time of the same SPCIES solvers when Algorithm \ref{Algorithm1} is not executed, i.e., without updating $A$, $B$, $Q$ and/or $R$.
\item Column ``\%" shows (Avg. update-time)/(Avg. update-time + Avg. solve-time).
    The ADMM penalty parameter of SPCIES is set to $\rho = 0.01$.
\end{tablenotes}
\caption{Computation times of the MPC solvers applied to the CSTR system for the MPC formulation \eqref{laxMPC}.}
\label{update_time_comparison}
\end{table*}

\begin{figure*}
    \centering
    \begin{subfigure}{0.47\textwidth}
        \includegraphics[width=\linewidth]{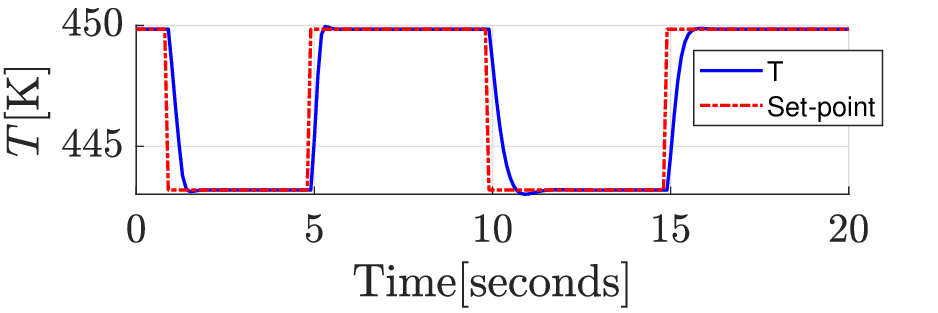}
        \caption{State $T$.}
        \label{state3}
    \end{subfigure}%
    \hfill
    \begin{subfigure}{0.47\textwidth}
        \includegraphics[width=\linewidth]{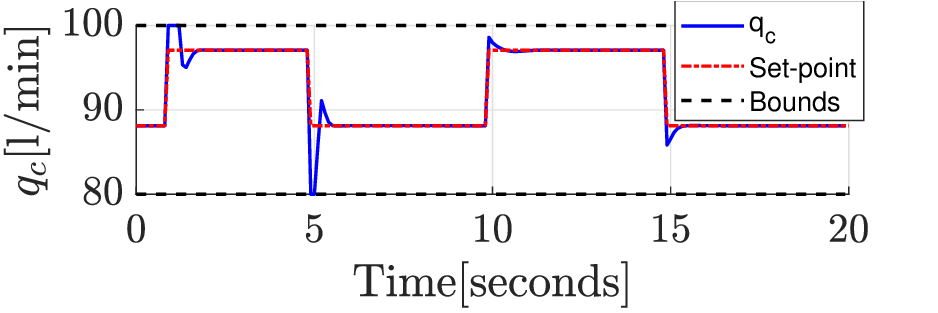}
        \caption{Input $q_c$.}
        \label{input2}
    \end{subfigure}%

    \caption{Non-linear CSTR model controlled with LPV frozen scheduling parameter MPC approach.}
    \label{experiment}
\end{figure*}

\begin{figure*}
    \centering
    \begin{subfigure}{0.33\textwidth} \label{subfig:nx_dependency}
        \includegraphics[width=\linewidth]{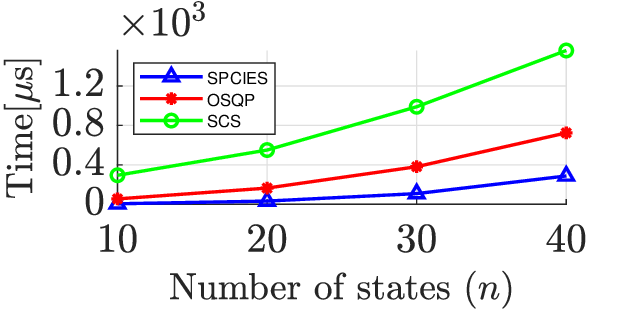}
        \caption{Dependency on $n$ ($m=2, \ N=5$)}
    \end{subfigure}%
    \begin{subfigure}{0.33\textwidth}
        \includegraphics[width=\linewidth]{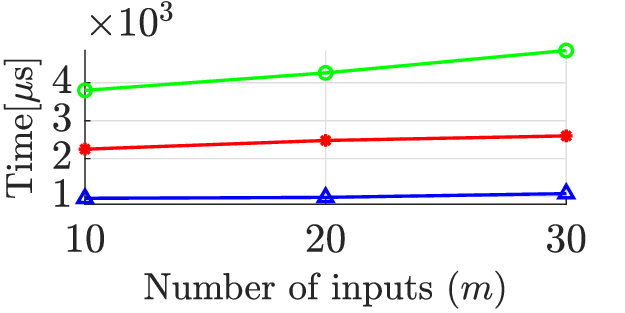}
        \caption{Dependency on $m$ ($n=60, \ N=5$)}
        \label{subfig:nu_dependency}
    \end{subfigure}%
    \begin{subfigure}{0.34\textwidth}
        \includegraphics[width=\linewidth]{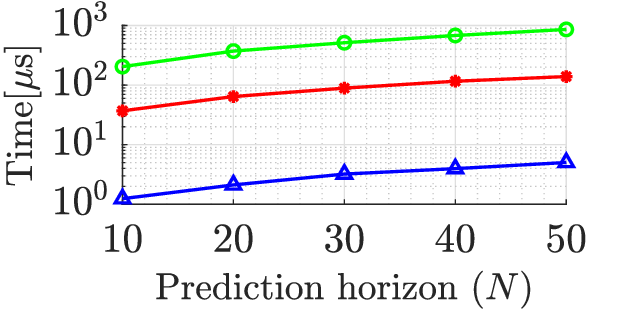}
        \caption{Dependency on $N$ ($n=4, \ m=2$)}
        \label{subfig:N_dependency}
    \end{subfigure}%
    \caption{Dependency of update times on $n$, $m$ and $N$.}
    \label{fig:update_dependency}
\end{figure*}

We consider the continuous stirred-tank reactor (CSTR) system from \cite{9505451}, where a reactant $A$ is transformed into a product $B$ by an exothermic and irreversible chemical reaction.
The objective is to control the reaction volume ($V$), the concentration of $A$ ($C_A$), and the temperature inside the reactor ($T$).
The manipulable variables are the output flow rate ($q_s$) and the input coolant flow rate ($q_c$).
The non-linear continuous-time equations of the system are given by
\begin{subequations}\label{LPV_model}
    \begin{align*}
        \frac{dV(t)}{dt} &= q_e-q_s(t), \\
        \frac{dC_A(t)}{dt} &= \frac{q_e}{V(t)}(C_{Ae}-C_A(t))-k_0 e^{\frac{-E}{RT(t)}}C_A(t),\\
        \frac{dT(t)}{dt} &= \frac{q_e}{V(t)}(T_e-T(t))-k_1e^{\frac{-E}{T(t)}}C_A(t)\\
        &\quad+ \frac{q_c(t)}{V(t)}k_2(1-e^{\frac{-k_3}{q_c(t)}})(T_{ce}-T(t)), \nonumber
    \end{align*}
\end{subequations}
where $k_1 = \frac{\Delta H k_0}{\rho C_p}$, $k_2 = \frac{\rho C_{pc}}{\rho_c C_p}$, $k_3 = \frac{h_A}{\rho_c C_{pc}}$ and the values of the system parameters are presented in \cite{9505451}. 
We consider an LPV model of the system, obtained from the above non-linear equations as shown in \cite{9505451}, with a frozen scheduling parameter, i.e., the matrices of the discrete LPV system model change at every sample time, and thus also the MPC prediction model, which remains constant along its prediction horizon.

We take the sampling time of the system as $0.1$ seconds, and control it using the MPC formulation \eqref{laxMPC}.
We implement the ADMM and FISTA solvers from \cite{Krupa_TCST_21}, available in the SPCIES toolbox (version v0.3.8) \cite{krupa2020spcies}, which include Algorithm~\ref{Algorithm1}.
We also solve the MPC optimization problem using the OSQP (version 0.6.2) \cite{osqp} and SCS (version 3.2.3) \cite{scs} solvers, which include a generic update procedure, as discussed in Remark~\ref{rem:generic}.
We note that OSQP also implements ADMM and SCS implements a similar operator splitting algorithm.
The simulations are run on an Intel Core i5-1135G7.

Table \ref{update_time_comparison} shows the computation times when taking average results from $200$ simulations using random initial conditions and references within the system constraints described in~\cite{9505451}.
Computation times are separated into ``update-time", which is the time taken by the procedure to update the solver ingredients due to the modifications of the MPC parameters and ingredients (this includes the computation time of Algorithm~\ref{Algorithm1} in the case of the SPCIES solvers), and ``solve-time", which is the remaining computation time.
The results show a clear difference in terms of computation time between the update-time of Algorithm~\ref{Algorithm1} w.r.t. OSQP or SCS. 

In order to show that the updating procedure we propose works properly when combined with time-varying MPC approaches, and that the time results we shown in Table~\ref{update_time_comparison} are obtained from a suitably designed MPC controller (in the sense of proper closed-loop behavior), we include Figure~\ref{experiment}, which shows the closed-loop result of one of the simulations of Table~\ref{update_time_comparison} using the solver from \cite{krupa2020spcies}, where the MPC model changes every sample time due to the LPV approach used, and the cost function matrices $Q$ and $R$ were changed from $Q= 10 I_n$ to $Q= I_n$ and from $R= \diag(10, 0.1)$ to $R= I_m$ at time $t = 9$ seconds, explaining the different system behavior between the first and second half of the simulation

Finally, Figure~\ref{fig:update_dependency} presents results on the computation time of Algorithm \ref{Algorithm1} in terms of the number of states ($n$), inputs ($m$) and the prediction horizon ($N$) as well as the update times when using the OSQP and SCS solvers. Note that Figure~\ref{subfig:N_dependency} is shown in logarithmic scale in order to improve its readability.
The results show that the proposed approach provides faster computation times than when using state-of-the-art generic QP solvers, as well as better scaling as the dimensions of the MPC problem increase.
This is due to our particularization of the update procedure to the specific matrix structure that arises in linear MPC problems.
We note that the update-time of Algorithm~\ref{Algorithm1} grows linearly with the prediction horizon $N$, as discussed in Section~\ref{sec:online}, since the number of matrices $\alpha^k$ and $\beta^k$ is linear in $N$.

\begin{remark}
We do not include a comparison with \cite{6571216} because we focus on comparisons with state-of-the-art solvers available online.
The computational complexity of the ``update-step" \cite[Algorithm 3]{6571216} does not depend on the prediction horizon $N$ if the terminal cost of the MPC controller is taken as the solution of the discrete LQR Riccati equation.
In this case, its computation time would be smaller than the one obtained with Algorithm~\ref{Algorithm1}.
However, if this is not the case, or if the system is not LTI along the prediction horizon, Algorithm~\ref{Algorithm1} outperforms it in terms of the number of required operations.
\end{remark}

\section{Conclusion} \label{sec:conclusions}

This article has presented an efficient method to perform the matrix factorization required by several FOM optimization solvers from the literature for linear MPC formulations.
The proposed update method is based on exploiting the particular structure that typically arises from standard MPC formulations, thus providing a computational advantage w.r.t. the general sparse update procedures used by many state-of-the-art QP solvers, as shown in the numerical results we have presented, where computation times were up to an order of magnitude faster.
Combined with the solvers presented in \cite{Krupa_TCST_21} (available in \cite{krupa2020spcies}), which are also based on exploiting MPC-specific matrix structures, this becomes an interesting option for applying MPC in under-powered devices when an online change of the prediction model or the tuning parameters of the MPC controller are required.

\begin{appendix}

This section provides an intuitive overview of how the update expressions for $\alpha^k$ and $\beta^k$ presented in Section~\ref{sec:online} are derived.
Recalling the definition \eqref{shortcuts}, $W \doteq G(H + I_\rho)^{-1}G\T$ is given by
\begin{equation} \label{W}
        W =
        \begin{bmatrix}
            Z + \hat{Q} & -\hat{Q}A^{\top} & 0 & \cdots & 0\\
            -A\hat{Q} & X_Q & -\hat{Q}A^{\top} & 0 & 0\\
            0 & \ddots & \ddots & \ddots & 0 \\
            0 & 0 & -A\hat{Q} & X_Q & -\hat{Q}A^{\top} \\
            0 & 0 & 0 & -A\hat{Q} & X_T
        \end{bmatrix}.
\end{equation}
The structure of its Cholesky decomposition is given by \eqref{alpha_beta}.
Using the identity $W = W_c\T W_c$, we can derive the following equations relating $\alpha^k$ and $\beta^k$ with the matrices in \eqref{W}:
\begin{align*}
    &(\beta^1)\T \beta^1 = Z + \hat{Q}, \; (\beta^1)\T \alpha^1 = -(A\hat{Q})\T,\\
    &(\beta^2)\T \beta^2 =  X_Q-(\alpha^1)\T \alpha^1, \; (\beta^2)\T \alpha^2 = -(A\hat{Q})\T,\\
    &\vdots \\
    & (\beta^N)\T \beta^N =  X_T-(\alpha^{N-1})^\top \alpha^{N-1}.&
\end{align*}
For matrices $\beta^k$, we can use their upper-triangular nature to further expand the above equations to obtain expressions for their non-zero elements.
For instance, for $\beta^1$, we have
\begin{align*}
    & (\beta^1)^\top \beta^1 {=}
    \begin{bmatrix}
        \beta^1_{1,1} & 0 & 0 \\
        \vdots & \ddots & 0 \\
        \beta^1_{1,n} & \cdots & \beta^1_{n,n}
    \end{bmatrix}
    \begin{bmatrix}
        \beta^1_{1,1} & \cdots & \beta^1_{1,n}\\
        0 & \ddots & \vdots\\
        0 & 0 & \beta^1_{n,n}
    \end{bmatrix}
    {=} Z {+} \hat{Q},			
\end{align*}
from where we derive
\begin{equation} \label{eq:beta_1}
\begin{aligned}
    &(\beta^1_{1,1})^2 = (Z + \hat{Q})_{1,1}, \ \beta^1_{1,2} = (Z + \hat{Q})_{1,2}/\beta^1_{1,1}, \ \dots ,\\
    & \beta^1_{1,n} = (Z + \hat{Q})_{1,n}/\beta^1_{1,1}, \    (\beta^1_{2,2})^2 = (Z + \hat{Q})_{2,2} - (\beta^1_{1,2})^2,\\
    &\beta^1_{2,3} = [(Z + \hat{Q})_{2,3} - \beta^1_{1,2} \beta^1_{1,3}]/\beta^1_{2,2}, \ \dots ,\\
    &\beta^1_{2,n} = [(Z + \hat{Q})_{2,n} - \beta^1_{1,2} \beta^1_{1,n}]/\beta^1_{2,2}, \ \dots , \\
    &(\beta^1_{n,n})^2 = (Z + \hat{Q})_{n,n} - (\beta^1_{1,n})^2 - \cdots - (\beta^1_{n-1,n})^2.	
\end{aligned}
\end{equation}
Continuing with $\beta^2$, we have
\begin{equation*}
        (\beta^2)^\top \beta^2 = X_Q -(\alpha^1)^\top \alpha^1,
\end{equation*}
which leads to the same expressions as \eqref{eq:beta_1} but substituting $\beta^1$ with $\beta^2$ and $Z + \hat{Q}$ with $X_Q -(\alpha^1)^\top \alpha^1$.
The same procedure can be performed for each $\beta^k$ up to $\beta^N$, which satisfies
\begin{equation*}
    (\beta^N)\T \beta^N = X_T-(\alpha^{N-1})^\top \alpha^{N-1}.	
\end{equation*}
Equations \eqref{beta:eq} and \eqref{beta:ineq} are a generalization of \eqref{eq:beta_1}, where $\Gamma^k$ \eqref{Gamma_expressions} accounts for the matrices of \eqref{W} that affect each $\beta^k$, e.g., $Z + \hat{Q}$ for $\beta^1$ or $X_T$ for $\beta^N$, and $\gamma$ \eqref{gamma_expression} accounts for the effect of the ``previous" values of the elements of $\beta^k$ and $\alpha^{k-1}$ that affect the current $\beta^k_{i, j}$, as shown in \eqref{eq:beta_1}.
Finally, we take the convention $\alpha^0 = 0$ so that we can also add $\alpha^0$ to the expression for $\beta^1$ in \eqref{lax_formulas}.

\end{appendix}

\bibliographystyle{IEEEtran}
\bibliography{IEEEabrv,mybib}

\end{document}